\documentclass[12pt]{article}

\usepackage{mathrsfs}
\usepackage{latexsym}  
\usepackage{amssymb}
\usepackage{graphicx}
\usepackage{amsmath}
\usepackage{ulem}

\newcommand{\be}{\begin{equation}}
\newcommand{\ee}{\end{equation}}
\newcommand{\bea}{\begin{eqnarray}}
\newcommand{\eea}{\end{eqnarray}}
\newcommand{\bref}[1]{(\ref{#1})}
\usepackage{color}

\begin{document}
\begin{titlepage}
PACS numbers: 13.15.+g, 12.10.-g, 14.60.-z

\begin{center}
{\Large\bf  
Alternative Renormalizable 
$SO(10)$ GUTs \\ and Data Fitting	
}
\end{center}

\begin{center}

\vspace{0.1cm}

{\large Takeshi Fukuyama$^{a,}$%
\footnote{E-mail: fukuyama@se.ritsumei.ac.jp}},
{\large Nobuchika Okada$^{b,}$%
\footnote{E-mail: okadan@ua.edu}}
and
{\large Hieu Minh Tran$^{c,}$%
\footnote{E-mail: hieu.tranminh@hust.edu.vn}}

\vspace{0.2cm}

${}^{a} ${\small \it Research Center for Nuclear Physics (RCNP),
Osaka University, \\Ibaraki, Osaka, 567-0047, Japan}\\

${}^{b}${\small \it Department of Physics and Astronomy, University of Alabama, \\Tuscaloosa, Alabama 35487, USA} \\

${}^{c}${\small \it Hanoi University of Science and Technology, 1 Dai Co Viet Road, Hanoi, Vietnam}

\end{center}

\begin{abstract}

The alternative renormalizable minimal $SO(10)$ model is composed of the Yukawa couplings with $\textbf{10} \oplus \textbf{120}$ Higgs fields, whereas the right-handed (RH) neutrino Majorana masses are generated via the Witten mechanism. The gauge coupling unification is achieved together with a unique pattern of the fermion masses and mixing at the grand unification scale due to additional contributions of vector-like quarks to the standard model renormalization group equations. We perform the fitting of the model to the experimental data of charged fermion masses and the CKM matrix. The best fit point is obtained with large pulls for $m_c$, $m_s$, $m_b$, and $m_\tau$. For the modifications to the minimal model by adding either $\textbf{10}'$ or $\textbf{120}'$, a large deviation for the tau mass rules out all these models. In the case with the bottom and vector-like quark mixing, the mass matrices are well fitted the charged fermions but the bound on the light neutrino mass scale excludes this scenario. To ameliorate this deficit, we consider the two-step symmetry breaking scenario, $SO(10) \rightarrow SU(5) \rightarrow SU(3)_C \times SU(2)_L \times U(1)_Y$, with the $SO(10)$ breaking at the Planck scale leading to the radiatively generated RH neutrino Majorana masses being at the ordinary seesaw scale. For this case, we find the best fit point with $\chi^2 = 7.8$ consistent with experimental results including the neutrino sector. The largest deviation is 2.3$\sigma$ corresponding to the strange quark mass. Hence, a more precise determination of the strange quark mass can test this model. For the best fit point, we find the effective Majorana neutrino mass of $m_{\beta\beta} = 0.22$ meV and the sum of light neutrino masses as $\Sigma = 0.078$ eV, which are consistent with the current constraints from the search for the neutrinoless double beta decay and the CMB anisotropy measurement.

\end{abstract}
\end{titlepage}


\section{Introduction}

The grand unified theory (GUT) with the underlined symmetry of an $SO(10)$ group is a beautiful and convincing picture for the unification of interactions beyond the standard model (SM).
It is not only the unification of the SM gauge symmetries, but also that of the SM fermions of each generation into a single $\textbf{16}$-dimensional multiplet.
In the SM, it is well known that the gauge couplings do not unify at high energy scales.
This issue can be resolved in scenarios where physics at intermediate scales is introduced for the deformation of the renormalization group (RG) trajectories.
A popular direction motivated by the gauge hierarchy problem is the assumption of supersymetry (SUSY) with the SUSY breaking scale of about $\mathcal{O}(1-10)$ TeV.
The SUSY GUTs have been investigated in many aspects.

Since the fermion masses in $SO(10)$ models originate from the Yukawa couplings between Higgs fields and the tensor product of two matter multiplets
\cite{Slansky:1981yr},
\begin{eqnarray}
\textbf{16} \otimes \textbf{16}	&=&
	\textbf{10}_s \oplus \textbf{120}_a \oplus \textbf{126}_s \, ,
\end{eqnarray}
the construction of the Higgs sector can be varied.
In the minimal $SO(10)$ models with $\textbf{10} \oplus \overline{\textbf{126}}$ Higgs fields \cite{Review}, 
the heavy right-handed (RH) neutrinos obtain their Majorana masses from the vacuum expectation value (VEV) of $(\overline{\textbf{10}}, \textbf{1}, \textbf{3})$ under the subgroup $SU(4)_C\otimes SU(2)_L\otimes SU(2)_R~(G_{422})$ at the tree level, and the light left-handed (LH) neutrino masses are generated via a canonical seesaw mechanism \cite{seesaw}. 
This minimal model is supplemented with $\textbf{120}$ Higgs as a general renormalizable model which has served to correct minor mismatching with data in both SUSY and non-SUSY cases. 
So there arise interesting issues whether SUSY or non-SUSY and whether the minimal $SO(10)$ is the best $SO(10)$ model or the best GUT model in terms of reproducing experimental data.

If we respect the renormalizability and minimality, a model with $\textbf{10} \oplus {\textbf{120}}$ Higgs fields (we call this model the alternative minimal model) is alternative to the ordinary minimal model with $\textbf{10} \oplus \overline{\textbf{126}}$ Higgs fields. 
Unlike the case of $\overline{\textbf{126}}$,
$\textbf{120}$ Higgs has no VEV which directly generates the RH-neutrino Majorana (and LH-neutrino Majorana) masses.
The RH-neutrinos with vanishing masses at the tree level become massive by virtue of the Witten mechanism \cite{Witten}.
It gives an effective $\overline{{\bf 126}}$ coupling with $\textbf{16}$-plets of matters via quantum corrections at the 2-loop level. 
Thus, the seesaw scale in this scenario is relatively low because of the 2-loop suppression \cite{alternative, Bajc}.
Due to the non-renormalization theorem of SUSY theories, the Witten mechanism is peculiar and applied only in the non-SUSY framework.
Similar to the ordinary minimal $SO(10)$ model, the theoretical predictions of this model on the particle mass spectrum, the Cabibbo-Kobayashi-Maskawa (CKM) and Pontecorvo-Maki-Nakagawa-Sakata (PMNS) mixing matrices should be checked if they are all in agreement with experimental results.
Beside the minimal alternative $SO(10)$ model, we consider other two simple extensions with the Higgs sectors, respectively, comprised of $\textbf{10} \oplus \textbf{10}' \oplus \textbf{120}$ and 
$\textbf{10} \oplus \textbf{120} \oplus \textbf{120}'$.
Although the former is simpler in terms of $SO(10)$ representation, it has more degrees of freedom in the Yukawa couplings.
Meanwhile, the latter with larger $SO(10)$ representation is more predictive thanks to its smaller number of free inputs.
In all three considered models, we assume a single-scale symmetry breaking pattern,
$SO(10) \rightarrow 
	SU(3)_C \otimes SU(2)_L \otimes U(1)_Y$, 
 at the GUT scale ($M_G$) for simplicity.

Assuming the minimal SUSY extension of the SM (the MSSM) as an effective theory below the GUT scale,
the $SO(10)$ model with $\textbf{10} \oplus \textbf{120}$ Higgs fields was investigated in Refs. \cite{Matsuda:2000zp, JMP,Chang:2004pb,Lavoura:2006dv}. 
According to that, the result of data fitting without considering the neutrino sector was not satisfactory with much larger $\chi^2$ than the minimal model with $\textbf{10}+\overline{\bf 126}$ where $\chi^2\leq O(1)$ for SUSY \cite{Mimura, Babu} 
and $\chi^2 = 14.7$ for non-SUSY \cite{Ohlsson:2019sja} including the neutrino sector fitting.
Here, we are interested in the scenario where the successful gauge coupling unification is achieved by the contribution of additional vector-like quarks to the beta function coefficients \cite{Gogoladze:2010in}.
The vector-like quarks change the RGE trajectories of the Yukawa couplings compared to those in the SM 
and the MSSM, resulting in a distinctive pattern of fermion masses and mixing at the GUT scale.
We will also investigate the two-step symmetry breaking scenario where the GUT symmetry breaking chain $SO(10) \rightarrow SU(5) \rightarrow SU(3)_C \times SU(2)_L \times U(1)_Y$ happens at two different scales.
The matching of the alternative $SO(10)$ models to such 
an effective non-SUSY
 model at low energies has not been investigated so far, and will show 
new fitting results.
This analysis is essential to quantify how well the above model setups are compatible with the updated experimental results, for which the answer could not be derived from previous works.
The existence of vector-like quarks is also well motivated by the stability of the electroweak vacuum below the GUT scale \cite{Chen:2017rpn}, 
and could play the role in the observed experimental anomalies \cite{Cline:2017aed}.
In this paper, we employ the following procedure to fit the model parameters to the measured observables
including those in the neutrino sector.
Firstly, the experimental values and errors of the fermion masses and mixing parameters are evolved to the GUT scale, 
taking into account the effects of the additional vector-like quarks.
Using the $\chi^2$ function as a measure, we then look for the best fit points of these models.

The paper is organized as follows.
In the next section, the alternative minimal $SO(10)$ model is briefly reviewed, and its two simple extensions are introduced.
In Section 3, the experimental observables are evolved to the GUT scale by solving the RG equations.
In Section 4, we perform the data fittings for each alternative renormalizable $SO(10)$ models and show the results.
In Section 5, we introduce the mixing between the bottom and vector-like quarks mixing in the minimal alternative model and present the data fitting results for this scenario.
The last section is devoted to the conclusion.

\section{Alternative renormalizable $SO(10)$ models}

\subsection{The minimal model}

When an $SO(10)$ model includes the $\overline{\bf 126}$ Higgs field, 
   a VEV of the $(\overline{\bf 10},{\bf 1}, {\bf 3})$ component under $G_{422}$ generates the Majorana masses 
   for RH neutrinos ($N_R$'s) at the tree level. 
However, even without the $\overline{\bf 126}$ Higgs field, we can generate these Majorana masses 
   when the model includes a ${\bf 16}$-plet Higgs field ($H_{16}$), 
   since a bilinear product of $H_{16}^\dagger$ can play a role of the $\overline{\bf 126}$ Higgs field. 
An effective operator relevant to this mass generation is given by
\bea 
  {\cal L} \supset \frac{1}{M} {\bf 16}_i  {\bf 16}_j  H_{16}^\dagger  H_{16}^\dagger  ,
\eea  
where $M$ is a mass parameter. 
Although we cannot introduce such a higher dimensional term by hand in a renormalizable model, 
  it can be induced through quantum corrections at the 2-loop level as pointed out by Witten \cite{Witten}. 
This is very interesting since the loop corrections suppress the seesaw scale.

\begin{figure}
\begin{center}
\includegraphics[scale=0.172]{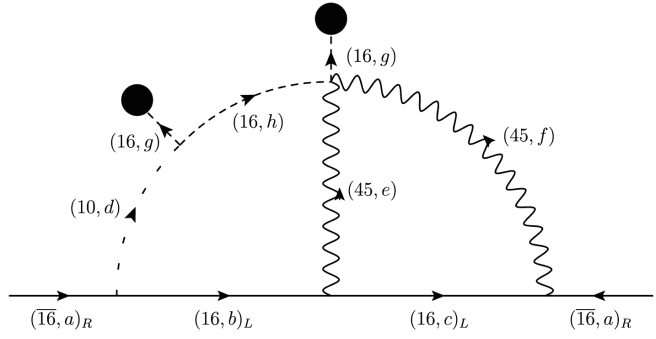}
\end{center}  
\caption{
The construction of the RH neutrino mass from
  ${\bf 10} \otimes {\bf 45} \otimes {\bf 45}$ via a 2-loop diagram. 
Shown in parenthesis are the $SO(10)$ and the broken subgroup 
 ($SU(5)$ or $G_{422}$) representations, which are summarized in  Table 2. 
Two blobs represent the insertion of $\langle H_{16} \rangle=M_G$.  
The crossed diagram should be added. 
}
\end{figure}

\begin{table*}
\begin{center}
\begin{math}
\begin{array}{|c|c|c|c|c|c|c|c|c|c|}
\hline
   & (a) & (b) & (c)  & (d) & (e) & (f)& (g)& (h) \\
\hline
SU(5) & \textbf{1} & \overline{\textbf{5}}& \textbf{10} & \textbf{5} & \textbf{10} & \textbf{10} & \textbf{1}& \textbf{5} \\
\hline
G_{422} & \overline{\textbf{4}},\textbf{1},\textbf{2} & \textbf{4},\textbf{2},\textbf{1} & \textbf{4},\textbf{2},\textbf{1} & \textbf{1},\textbf{2},\textbf{2}& \textbf{15},\textbf{1},\textbf{1}& \textbf{6},\textbf{2},\textbf{2}& \overline{\textbf{4}},\textbf{1},\textbf{2}& \textbf{4},\textbf{2},\textbf{1}\\
\hline
\end{array}
\end{math}
\end{center}
\caption{
Representations of particles in the 2-loop diagram under the $SO(10)$ subgroups. 
}
\end{table*}

In the simplest model discussed in Ref.~\cite{Witten}, the matter fermions couple directly  only to the ${\bf 10}$-plet Higgs and the $SO(10)$ gauge field of the ${\bf 45}$ representation. 
The basic idea is that ${\bf 126}$ representation is a 5th rank tensor 
  which can be constructed by the product ${\bf 10} \otimes {\bf 45} \otimes {\bf 45}$ 
  with a vector ${\bf 10}$ and a 2nd rank tensor ${\bf 45}$. 
In fact, the $N_R$ mass is generated by quantum corrections at the 2-loop level 
  as shown in Fig.~1, when $H_{16}$ develops its vacuum expectation value (VEV) $(\textbf{1},-5$) for the subgroup $SU(5)\otimes U(1)_X$ or a 
VEV $(\overline{\textbf{4}}, \textbf{1},\textbf{2})$  for the subgroup $SU(4)\otimes SU(2)_L\otimes SU(2)_R$%
.
Here, note that a triple scalar coupling among the Higgs fields of 
$\textbf{10}$-plet and 
$\textbf{16}$-plets also plays a crucial role: 
\bea 
  {\cal L} \supset \lambda_{10} M_G H_{10} H_{16} H_{16},   
\label{3-scalar}  
\eea
 where we have parametrize the triple scalar coupling with $M_G$ and 
 a dimensionless coupling constant $\lambda_{10}$. 
The resultant $N_R$ mass is estimated as  
\bea
M_R = \left( \frac{m_q}{M_W} \right) \epsilon_{10} \left( \frac{\alpha_G}{\pi} \right)^2 M_G.
\label{MR}
\eea
Here, we have used a relation,  $Y_{10} \sim m_q/M_W$
   between the Yukawa coupling of $H_{10}$ and an up-type quark mass $m_q$, 
   and $\epsilon_{10}$ represents a mixing angle between $H_{10}$ and $H_{16}$ 
   induced by their coupling in Eq.~(\ref{3-scalar}) with a VEV of $H_{16}$.    
Note that the $M_R$ scale is much lower than the usual seesaw scale of the model $\sim M_G$.    
In the present $SO(10)$ model with only one Yukawa coupling $Y_{10}$, 
  the Dirac neutrino mass matrix is the same as the up-type quark mass matrix. 
Therefore, the light neutrino mass $m_{\nu_L}$ due to the type I seesaw mechanism,
\be
m_{\nu_L}=-M_D^TM_R^{-1}M_D,
\ee
is given by
\bea
m_{\nu_L} = m_q  \left[ \epsilon_{10} \left( \frac{\alpha_G}{\pi} \right)^2 \right]^{-1} \frac{M_W}{M_G}.
\label{mnu}
\eea
As in Ref.~\cite{Witten}, we estimate $m_{\nu_L}=10^{-7} \, m_q$ 
  by using $(\alpha_G/\pi)^2 =10^{-5}$, $M_G=10^{15}$ GeV and $\epsilon_{10}=0.1$. 
Clearly, the light neutrino mass spectrum predicted by this formula is unrealistic. 
For example, the heaviest light neutrino mass is $10^{-7}\, m_t \sim 20$ keV, 
   where $m_t=173$ GeV is the top quark mass. 
Thus, the light neutrino masses from the type I seesaw with the 2-loop induced $M_R$ are too heavy. 
This is due to the quark mass $m_q$ insertions in Eqs. \bref{MR} and \bref{mnu}, which originate from the single Yukawa coupling $Y_{10}$. A simple way to ameliorate the problem is to add one more Yukawa coupling with ${\bf 120}$ Higgs.
${\bf 120}$ Higgs field includes two pairs of $SU(2)_L$ Higgs doublets $({\bf 15,~2,~2})$ and $({\bf 1,~2,~2})$ unlike the $\overline{\bf 126}$ Higgs field. 
Their general renormalizable mass formula was given in \cite{JMP}.
  ${\bf 120}$ Higgs does not involve $(\overline{{\bf 10}}, {\bf 3}, {\bf 1}) \oplus ({\bf 10}, {\bf 1}, {\bf 3})$ under the subgroup $G_{422}$. Hence, the Majorana neutrino masses are not generated at the tree-level but via the two-loop correction a la Witten mechanism.  
With the ${\bf 120}$ Higgs field, the Yukawa interactions are given by    
\bea
 {\cal L}_Y = Y_{10}^{ij} {\bf 16}_i {\bf 10}_H {\bf 16}_j 
           +Y_{120}^{ij} {\bf 16}_i {\bf 120}_H {\bf 16}_j.  
\label{Y_120}
\eea
With the  VEVs of three pairs of Higgs doublets (one in ${\bf 10}_H$ and two in ${\bf 120}_H$), 
  the fermion mass matrices at $M_G$ are described as \cite{JMP}
\bea
\label{MudDe}
M_u &=& c_{10}\, M_{10} + c_{120} \, m_{120} 
+ c'_{120} \,m'_{120}, 
\\
M_d &=& M_{10} + m_{120} + m'_{120}, 
\\
M_D &=& c_{10} \,M_{10} + c_{120} \,m_{120} 
- 3 \,c'_{120} \,m'_{120}, 
\\
M_e &=& M_{10} + m_{120} - 3 \,m'_{120} , 
\eea
where $M_u$, $M_d$, $M_D$, and $M_e$ denote the up-type quark,
down-type quark, Dirac neutrino, and charged-lepton, respectively.
Here, the mass matrices 
$m_{120}$ and $m'_{120}$ come from  
$({\bf 1,~2,~2})$ and $({\bf 15,~2,~2})$ of ${\bf 120}$, respectively. The mass matrices of the right-hand sides are defined as
$M_{10}= Y_{10}\, \alpha_d^1 v \cos\beta$,
$m_{120}= Y_{120}\, \alpha_d^2 v \cos\beta$, 
and 
$m'_{120}= Y_{120}\, \alpha_d^3 v \cos\beta$
respectively, and the coefficients are defined as
$c_{10}= (\alpha_u^1/\alpha_d^1) \tan \beta$,
$c_{120}= (\alpha_u^2/\alpha_d^2) \tan \beta$, $c'_{120} = (\alpha_u^3/\alpha_d^3) \tan \beta$. 
$\alpha_{u,d}^i$ are the complex elements of the unitary matrices which make the light pair of Higgs doublets $H_u$ and $H_d$ from many doublets. These complex values depend on the Higgs potential. See \cite{JMP} for details. Here it is sufficient to recognize that the coefficients $c$s are independent complex numbers.
These mass matrices are directly connected with low-energy
observations and are resumed as
\begin{eqnarray}
M_u &=& c_{10} M_{10} + c_{120}^u M_{120}, \label{mu1}\\
M_d &=&     M_{10} +     M_{120},  \label{md1}	\\ 
M_D &=& c_{10} M_{10} + c_{120}^n M_{120},	\label{mD1}\\
M_e &=&     M_{10} + c_{120}^e M_{120}.	\label{me1}
\end{eqnarray}
Here, $M_{120}$ is
\begin{equation}
M_{120} = m_{120} + m'_{120} \, ,
\end{equation}
and the complex coefficients $c_{120}^{u,n,e}$ are 
\begin{eqnarray}
c_{120}^u 	&=&	
	\frac{\alpha_u^2+\alpha_u^3}	{\alpha_d^2+\alpha_d^3}\tan\beta, \quad \\
c_{120}^n	&=&
	\frac{\alpha_u^2-3\alpha_u^3}{\alpha_d^2+\alpha_d^3}\tan\beta, \quad \\
c_{120}^e	&=&
	\frac{\alpha_d^2-3\alpha_d^3}{\alpha_d^2+\alpha_d^3}\tan\beta  .
\end{eqnarray}

The mass matrices $M_{10}$ and $M_{120}$ are respectively complex symmetric and antisymmetric. 
Since the neutrino Dirac mass matrix $M_D$ is not equal to $M_u$,
hence the problem appeared in Eq. \bref{mnu} does not occur.
Here, the light LH neutrino mass can be estimated as
\begin{eqnarray}
m_{\nu_L} & \sim &
	\frac{(M_u + (c^n_{120} - c^u_{120}) M_{120})^2}{M_u - c^u_{120} M_{120}} \times
	\left[ \epsilon_{10} \left( \frac{\alpha_G}{\pi} \right)^2 \right]^{-1} \frac{M_W}{M_G}.
\end{eqnarray}
With appropriate values of the parameters $c^n_{120}$, $c^u_{120}$, and $M_{120}$, the light LH neutrino mass can be of the right order.
For example, in the case $c^u_{120} M_{120} = \mathcal{O}(M_u)$, we can set $c^n_{120} M_{120}$ so as to almost cancel the numerator such that 
the light LH neutrino mass can be $m_{\nu_L} = \mathcal{O}(0.1)$ eV. 
In this estimation, the hierarchy in $m_{\nu_L}$ is similar to that in $M_u$.

Given the experimental data at that time, the system with ${\bf 10} \oplus {\bf 120}$ Higgs fields was shown to be consistent with the realistic charged fermion mass spectra \cite{Matsuda:2000zp} when the Yukawa coupling matrices of $\textbf{10}$ and $\textbf{120}$ are respectively assumed to be real and pure imaginary for simplicity. 
The fitting was not exhausted with full data.
Afterward, more elaborate data fittings of this model have been performed in the Ref. \cite{Chang:2004pb,Lavoura:2006dv} for the SUSY case.  We perform the data fitting of our models in Section 4.

\subsection{Two simple extensions}

Beside the alternative minimal $SO(10)$ model, we also consider some extensions of it.
As simple possibilities, we introduce one more Higgs multiplet of either $\textbf{10}'$-plet or $\textbf{120}'$-plet.
Although the $\textbf{10}'$-plet is simpler than the $\textbf{120}'$-plet in terms of the field content, the Yukawa sector of the former case has more independent parameters than that of the latter one.


In the extension with the Higgs sector $\textbf{10} \oplus \textbf{120} \oplus \textbf{120}'$, the Yukawa sector is given as
\begin{eqnarray}
\mathcal{L}_Y^1	&=&
	Y_{10}^{ij} \textbf{16}_i \textbf{10}_H \textbf{16}_j +
	Y_{120}^{ij} \textbf{16}_i \textbf{120}_H \textbf{16}_j +  			Y_{120'}^{ij} \textbf{16}_i \textbf{120}'_H \textbf{16}_j .
\end{eqnarray} 
There are five pairs of Higgs doublets developing VEVs in this model (one in $\textbf{10}_H$, and two in each representation of $\textbf{120}_H$ and $\textbf{120}'_H$) where only one pair of their linear combination ($H_u, H_d$) are assumed to be light while the other four are heavy \cite{Chang:2004pb}. 
Once the light Higgs doublets develop their VEVs ($v_u, v_d$), the SM fermion masses are generated via the Yukawa couplings at $M_G$:
\begin{eqnarray}
M_u	&=&	c_{10} M_{10} + c_{120}^u m_{120} + {c'}_{120}^u m'_{120} ,	\\
M_d	&=&	M_{10} + m_{120} + m'_{120} ,	\\
M_D	&=&	c_{10} M_{10} + c_{120}^n m_{120} + {c'}_{120}^n m'_{120} ,	\\
M_e	&=&	M_{10} + c_{120}^e m_{120} + {c'}_{120}^e m'_{120} .	
\end{eqnarray}
After a re-parameterization,
\begin{eqnarray}
M_{120}	&=&	m_{120} + m'_{120} ,	\\
M'_{120}
		&=&	c_{120}^u m_{120} + {c'}_{120}^u m'_{120} ,	\\
d_{120}^n &=&
	\frac{c_{120}^n {c'}_{120}^u - {c'}_{120}^n c_{120}^u}{{c'}_{120}^u - c_{120}^u} ,	\\
{d'}_{120}^n &=&
	\frac{{c'}_{120}^n - c_{120}^n }{{c'}_{120}^u - c_{120}^u} ,	\\
d_{120}^e &=&
	\frac{c_{120}^e {c'}_{120}^u - {c'}_{120}^e c_{120}^u}{{c'}_{120}^u - c_{120}^u} ,	\\
{d'}_{120}^e &=&
	\frac{{c'}_{120}^e - c_{120}^e }{{c'}_{120}^u - c_{120}^u} ,
\end{eqnarray}
the fermion mass matrices at the GUT scale in this model are determined by the GUT parameters as
\begin{eqnarray}
M_u	&=&	c_{10} M_{10} + M'_{120} ,	\\
M_d	&=&	M_{10} + M_{120} ,	\label{md2}\\
M_D	&=&	c_{10} M_{10} + d_{120}^n M_{120} + {d'}_{120}^n M'_{120} ,	\\
M_e	&=&	M_{10} + d_{120}^e M_{120} + {d'}_{120}^e M'_{120} ,	\label{me2}
\end{eqnarray}
where the coefficients 
$c_{10}$, 
$d_{120}^n$, ${d'}_{120}^n$, 
$d_{120}^e$, and ${d'}_{120}^e$ are complex. 
The matrix $M_{10}$ is complex symmetric, and
the matrices $M_{120}$ and $M'_{120}$ are both  complex antisymmetric.


In the other extension, the Higgs sector consists of 
$\textbf{10} \oplus \textbf{10}' \oplus \textbf{120}$ representations.
The Yukawa sector is given as
\begin{eqnarray}
\mathcal{L}_Y^2	&=&
	Y_{10}^{ij} \textbf{16}_i \textbf{10}_H \textbf{16}_j +
	Y_{10'}^{ij} \textbf{16}_i \textbf{10}'_H \textbf{16}_j + 
	Y_{120}^{ij} \textbf{16}_i \textbf{120}_H \textbf{16}_j .
\end{eqnarray}
Similar to the case of $\textbf{10} \oplus \textbf{120} \oplus \textbf{120}'$ Higgs fields,
in this case the fermion mass matrices at the GUT scale are determined by the GUT parameters  as
\begin{eqnarray}
M_u	&=&	M'_{10} + c_{120}^u M_{120} ,	\\
M_d	&=&	M_{10} + M_{120} ,	\label{md3}\\
M_D	&=&	M'_{10} + c_{120}^n  M_{120} ,	\\
M_e	&=&	M_{10} + c_{120}^e M_{120} , \label{me3}
\end{eqnarray}
where  the coefficients $c_{120}^{u,n,e}$ are complex, 
the matrices $M_{10}$ and $M'_{10}$ are both complex symmetric, and
the matrix $M_{120}$ is complex antisymmetric.


To fit the model parameters to the experimental data, we firstly evaluate the experimental values of observables at the GUT scale by solving the RG equations (RGEs).
The theoretical prediction of these observables at the GUT scale to be compared with the experimental values are determined for each point of the free parameter space.
Then, we scan over the parameter space to find the best fit point.

\section{Observables at the GUT scale}

In the SM, it is well-known that the three gauge couplings are not successfully unified at high energy scales.
To resolve this shortcoming, we assume that below the GUT scale there are two pairs of vector-like quarks 
($Q_L, Q_R, D_L, D_R$) carrying the SM charges \cite{Gogoladze:2010in} as given in Table \ref{vector-likeQ}.

\begin{table}[h]
\begin{center}
\begin{math}
\begin{array}{|c||c|c|c|}
\hline
		&	SU(3)_C			&	SU(2)_L	&	U(1)_Y	\\
\hline
\hline
Q_{L,R}		&	\textbf{3}		&	\textbf{2}		&	\frac{1}{6}		\\
D_{L,R}		&	\textbf{3}		&	\textbf{1}		&	\frac{1}{3}		\\
\hline
\end{array}
\end{math}
\end{center}
\caption{The SM charges of the additional vector-like quarks}
\label{vector-likeQ}
\end{table}

Similar to the case of a $SU(5)$ model
\cite{Okada:2017dqs}, in the framework of the $SO(10)$ GUT, these vector-like quarks are embedded in two pairs of vector-like $SO(10)$ representation,
$F_{16}^Q+\bar{F}_{16}^Q$ and 
$F_{16}^D+\bar{F}_{16}^D$.
The Lagrangian for them is written as follows:
\begin{eqnarray}
\mathcal{L}_{SO(10)} &\supset&
	\text{Tr}
	\left[
		\bar{F}_{16}^Q (Y_{16}^Q \Sigma - M_{16}^Q) F_{16}^Q
	\right] 	+ 
	\text{Tr}
	\left[
		\bar{F}_{16}^D (Y_{16}^D \Sigma - M_{16}^D) F_{16}^D
	\right] ,
\end{eqnarray}
where $Y_{16}^{Q,D}$ are Yukawa couplings, 
$M_{16}^{Q,D}$ are vector-like fermion masses, and
$\Sigma$ is an $SO(10)$ adjoint Higgs field of \textbf{45}-plet
whose vacuum expectation value with 
$\langle H_{16} \rangle$
breaks $SO(10)$ to $SU(3)_C \times SU(2)_L \times U(1)_Y$.
By tuning $Y_{16}^{Q,D}$ and $M_{16}^{Q,D}$ according to the method of generating doublet-triplet mass splitting, the vector-like quarks ($Q_L, Q_R, D_L, D_R$) can be as light as $\mathcal{O}$(TeV), while other components are super-heavy with masses around the GUT scale.

At low energies, the the Lagrangian of the effective theory includes that of the SM and the mass terms for the vector-like quarks:
\begin{eqnarray}
\mathcal{L}_{\text{mass}} &\supset &
	m_Q \bar{Q}_L Q_R + m_D \bar{D}_L D_R + \text{h.c.} \, .
\end{eqnarray}
In this analysis, we assume that 
$m_Q = m_D = M_F = \mathcal{O}$(TeV) for simplicity.

From the electroweak scale to the vector-like quark scale $M_F$, the gauge coupling running follows the SM RGEs.
At the 2-loop level, they are given by 
\cite{Machacek:1983tz}
\begin{eqnarray}
\dfrac{d g_i}{d \ln \mu}	&=&
	\frac{b_i}{16\pi^2} g_i^3 +
	\frac{g_i^3}{(16\pi^2)^2}
		\left(
		\sum_{j=1}^3 B_{ij} g_j^2 - 
		\text{Tr} (	C_{i,u} Y_u^\dagger Y_u +
					C_{i,d} Y_d^\dagger Y_d +
					C_{i,e} Y_e^\dagger Y_e)
		\right) ,
\end{eqnarray}
where the values of the index
$i = 1,2,3$ correspond to the SM gauge groups $U(1)_Y, SU(2)_L,$ and $SU(3)_C$.
The other coefficients are as follows
\begin{eqnarray}
b_i &=& \left( 	\frac{41}{10}, 
			-\frac{19}{6},
			- 7 
		\right), \qquad (i = 1,2,3),	\\
B_{ij}	&=&
	\left(
	\begin{array}{ccc}
	\frac{199}{50}	&	\frac{27}{10}	&	\frac{44}{5}	\\
	\frac{9}{10}	&	\frac{35}{6}	&	12	\\
	\frac{11}{10}	&	\frac{9}{2}		&	-26
	\end{array}
	\right)		,	\\
C_{i,f}	&=&
	\left(
	\begin{array}{ccc}
	\frac{17}{10}	&	\frac{1}{2}	&	\frac{3}{2}	\\
	\frac{3}{2}		&	\frac{3}{2}	&	\frac{1}{2}	\\
	2				&	2		&	0
	\end{array}
	\right)	,	\qquad (f = u,d,e).
\end{eqnarray}
With the existence of vector-like quarks between $M_F$ and the GUT scale  $M_G$, the beta function coefficients of the RGEs in this energy range for the SM gauge couplings are modified by additional contributions from these fermions 
\cite{{Gogoladze:2010in}}:
\begin{eqnarray}
b'_i &=& \left( 	\frac{2}{5}, 2,	2 \right), 
		\qquad (i = 1,2,3),	\\
B'_{ij}	&=&
	\left(
	\begin{array}{ccc}
	\frac{3}{50}	&	\frac{3}{10}	&	\frac{8}{5}	\\
	\frac{1}{10}	&	\frac{49}{2}	&	8	\\
	\frac{1}{5}		&	3		&	\frac{114}{3}
	\end{array}
	\right)		.
\end{eqnarray}
The RGEs for other parameters in this energy range are the same as those for the SM parameters. 
The vector-like quarks affect these RGEs indirectly via those of the gauge couplings.

\begin{figure}[h]
\begin{center}
\includegraphics[scale=0.98]{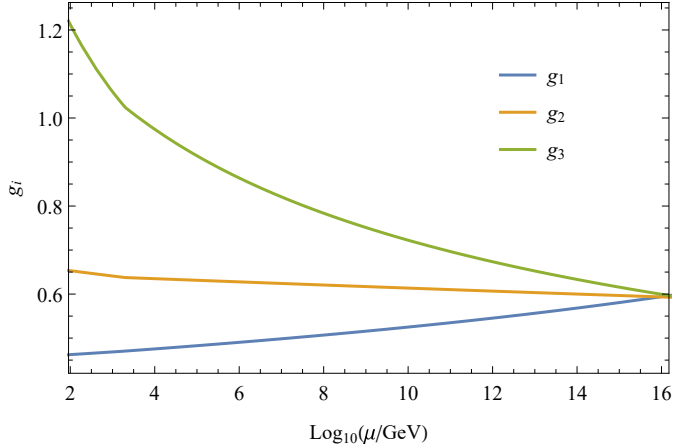}

\caption{
The RG runnings of three SM gauge couplings when the vector-like quark mass is $M_F = 2$ TeV.
In this case, the gauge couplings unify at $M_G = 1.5 \times 10^{16}$ GeV.
\label{unify}
}
\end{center}  
\end{figure}

By varying the vector-like quark mass scale, we find that a successful gauge coupling unification is achieved at 
$M_G = 1.5 \times 10^{16}$ GeV
for $M_F = 2$ TeV.
The RG evolution of three gauge couplings in this case is demonstrated in Figure \ref{unify}.

In order to evaluate the fermion masses and mixing at 
a high energy scale (like the GUT scale), 
all the relevant quantities must be determined at that scale for consistency \cite{Das:2000uk}. 
To determine the VEV at the GUT scale we used the RGE of the VEV in the paper by Arason \textit{et al} (Phys. Rev. D 46, 3945 (1992)) where the VEV running originates from the Higgs doublet’s wavefunction renormalization.%
\footnote{
Some of the previous works neglected the evolution of the VEV for simplicity, see for example Ref. \cite{Dueck:2013gca}. We thank the referee for commenting on this point.%
}
In our analysis, we solve the RGEs for the Yukawa coupling matrices and the VEV at the two-loop level \cite{Machacek:1983tz} from $m_Z$ to $M_G$
with the boundary conditions given at the electroweak (EW) scale.
The center values and the corresponding errors%
\footnote{
It is worth noting that observables with large errors (like $m_u$ and $m_d$) have negligible contributions to the RGEs, while the observable having significant contributions to the RGEs have a very small relative error (e.g $m_t$). 
We have checked that the correlations among errors are negligibly small and, in a good approximation, 
the upper/lower bound of an observable at the EW scale corresponds to its upper/lower (or lower/upper) bound at the GUT scale.
} 
of charged fermion masses and mixing at the EW scale are taken from Table 1 of Ref. \cite{Ohlsson:2018qpt} 
where the results in \cite{Xing:2007fb} were invoked. 
For the neutrino oscillation data at low energies, we use the values in Table 1 of  Ref. \cite{Esteban:2018azc}.

The charged fermion mass matrices at the GUT scale $M_G$ are calculated as
\begin{eqnarray}
M_f(M_G)	&=&	
	Y_f (M_G)
	\frac{v(M_G)}{\sqrt{2}},
	\qquad (f = u, d, e).
\label{mass}
\end{eqnarray}
The CKM mixing matrix at $M_G$ is determined from the rotation matrices that diagonalize the matrices 
$Y_u^\dagger Y_u$ and 
$Y_d^\dagger Y_d$ as follows:
\begin{eqnarray}
U_{\text{CKM}} (M_G)	&=&
	V_{uL}(M_G) \cdot
	V_{dL}^\dagger (M_G).
\label{CKM}
\end{eqnarray}

In Table \ref{GUTvalues}, we present the mean values and errors of the fermion masses and mixings at the GUT scale.
Here, $m_{u,c,t} ,$ $m_{d,s,b}$ and $m_{e,\mu,\tau}$ are, respectively, the singular values of the mass matrices $M_u, M_d,$ and $M_e$ in Eq. (\ref{mass}),
while $\lambda, A, \bar{\rho}, \bar{\eta}$ are the Wolfenstein parameters of the CKM mixing matrix in Eq. (\ref{CKM}).
$\Delta m_{21}^2$ and $\Delta m_{31}^2$ are the neutrino squared mass differences,
while $\theta_{12}$, $\theta_{23}$, $\theta_{13}$, and $\delta_\text{CP}$ are the parameters of the PMNS mixing matrix.

\begin{table}
\begin{center}
\begin{math}
\begin{array}{|c||c|c|}
\hline
\text{Observables}	&	\bar{x}_i &	\sigma_i	\\
\hline
\hline
m_u		&	0.40\times 10^{-3}	&	0.20\times 10^{-3}	\\
m_c		&	0.199				&	0.031	\\
m_t		&	60.4				&	1.7	\\
m_d		&	0.88\times 10^{-3}	&	0.46\times 10^{-3}	\\
m_s		&	0.018				&	0.006	\\
m_b		&	0.829				&	0.026	\\
m_e		&	0.440\times 10^{-3}	&	0.022\times 10^{-3}	\\
m_\mu 	&	0.093				&	0.005	\\
m_\tau	&	1.583				&	0.079	\\
\lambda	&	0.22469				&	0.39\times 10^{-3}	\\
A		&	0.932				&	0.012	\\
\bar{\rho}&	0.140				&	0.016	\\
\bar{\eta}&	0.356				&	0.010	\\
\Delta m_{21}^2	&	1.112\times 10^{-4}	&	0.032\times 10^{-4}	\\
\Delta m_{31}^2	&	3.798\times 10^{-3}	&	0.048\times 10^{-3}	\\
\theta_{12}	&	0.573	&	0.013	\\
\theta_{23}	&	0.867	&	0.017	\\
\theta_{13}	&	0.1503	&	0.0023	\\
\delta_\text{CP}	&	-2.50	&	0.59	\\
\hline
\end{array}
\end{math}
\end{center}
\caption{The charged fermion masses (in GeV), the Wolfenstein parameters of the CKM mixing matrix, 
the neutrino squared mass differences (in eV$^2$), and 
the parameters of the PMNS matrix at the GUT scale $M_G$.
The $\bar{x}_i$ and $\sigma_i$ columns correspond to the mean values and the uncertainties.}
\label{GUTvalues}
\end{table}

\section{Data fitting for alternative models}

To measure how well a point of the parameter space can predict the experimental data,
we use the $\chi^2$ function defined as
\begin{eqnarray}
\chi^2	&=& \sum_i p_i^2 =
	\sum_i
		\left( 
		\frac{x_i - \bar{x}_i}{\sigma_i}
		\right)^2	,
\end{eqnarray}
where $x_i$ indicate the theoretical values of the observables, 
$\bar{x}_i$ and $\sigma_i$ are the mean value and the uncertainties of the observables, 
and $p_i$ are the pulls to the corresponding observables. 
The sum is taken over all 13 observables in Table \ref{GUTvalues}.
The best fit point of the model that is the global minimum of the $\chi^2$ function
is found by performing two sequential steps.
Firstly, we randomly scan over the parameter space with the package MultiNest version 3.11 
\cite{Feroz:2007kg}
where the nested sampling algorithm is employed.
The ranges of the free inputs are chosen to be 
[-100,100] for those relevant to the coefficients ($c_{10}, c_{120}, d_{120}$), and
[$10^{-20},100$] (GeV) for those relevant to the mass matrices ($M_{10}, M_{120}$, $M'_{10}, M'_{120}$) with positive and negative signs allowed.
This scan is carried out many times to avoid falling into some local minimum.
Secondly, the smallest minimum found in the first step is refined further by using the Nelder-Mead simplex algorithm 
\cite{NelderMead} with a high precision.

\subsection{The minimal model with $\textbf{10} \oplus \textbf{120}$ Higgs multiplets}

In the $SO(10)$ model with 
$\textbf{10} \oplus \textbf{120}$ Higgs multiplets, we consider a general case where
the coefficients ($c_{120}^{u,n,e}$) and the matrices ($M_{10}$, $M_{120}$) in Eqs. (\ref{mu1})-(\ref{me1}) are complex.
Inherited from the properties of the corresponding Yukawa couplings, the mass matrices $M_{10}$ and $M_{120}$ are symmetric and antisymmetric, respectively.
With an appropriate choice of the $SO(10)$ fermion basis, we can set $M_{10}$ to be real and diagonal without loss of generality.
By rephasing u-type quarks, $c_{10}$ can be real.
Thus, relevant to the data fitting for the charged fermions masses and mixing there are totally 14 free real parameters including
four parameters for the complex coefficients $c_{120}^{u,e}$,
one parameter for the real coefficient $c_{10}$, 
three parameters for the real diagonal matrix $M_{10}$,
and six parameters for the complex antisymmetric matrix $M_{120}$.


The best fit point that we have found in the numerical analysis is
\begin{eqnarray}
c_{10}	&=&	67.126,		\\
c_{120}^u	&=&
	146.235	+ 64.985 i,	\\
c_{120}^e	&=&
	9.8535	+ 5.3155 i,	\\
M_{10}	&=&
	\left(
	\begin{array}{ccc}
	0.000516765	&	0	&	0	\\
	0	&	0.00432434	&	0	\\
	0	&	0	&	0.893733		
	\end{array}
	\right),	
\end{eqnarray}
\begin{eqnarray}
&& M_{120}	=	\nonumber	\\	 
&&	\left(
	\begin{array}{ccc}
	0	&	0.00034554 +0.00042528 i	&	0.0061402 -0.0059386 i 	\\
	-0.00034554 -0.00042528 i	&	0	&	0.013011 +0.023147 i	\\
	-0.0061402 +0.0059386 i	&	-0.013011 -0.023147 i	&	0		
	\end{array}
	\right).	 \nonumber	\\
\end{eqnarray}
In Table \ref{bestfit10+120}, the predicted values and the pulls of the charged fermion masses and the Wolfenstein parameters of the CKM mixing matrix at the GUT scale for the best fit point are presented.
We see that the total $\chi^2$ value is 69.5079 which is not good.
It is mostly due the large pulls toward opposite sides of the mean values for the bottom and tau masses:
\begin{eqnarray}
m_b^{\text{best-fit}} &=& \overline{m}_b + 2.5 \, \sigma_b ,	\\
m_\tau^{\text{best-fit}} &=& \overline{m}_\tau - 7.3 \, \sigma_\tau .
\end{eqnarray}
We can also see that there is a mild pull for the strange quark mass:
\begin{eqnarray}
m_s^{\text{best-fit}} &=& \overline{m}_s - 2.5 \, \sigma_s .	\label{pullms}
\end{eqnarray}

\begin{table}
\begin{center}
\begin{math}
\begin{array}{|c||c|c|}
\hline
\text{Observables}	&	x_i &	p_i	\\
\hline
\hline
m_u		&	0.40089\times 10^{-3}	&	0.0096	\\
m_c		&	0.20216				&	0.096	\\
m_t		&	60.3245				&	0.018	\\
m_d		&	0.49026\times 10^{-3}	&	0.87	\\
m_s		&	0.004064				&	2.5	\\
m_b		&	0.89461				&	2.5	\\
m_e		&	0.44180\times 10^{-3}	&	0.061	\\
m_\mu 	&	0.095153				&	0.43	\\
m_\tau	&	0.99239				&	7.4	\\
\lambda	&	0.22469				&	0.0055	\\
A		&	0.93283				&	0.031	\\
\bar{\rho}&	0.13985				&	0.020	\\
\bar{\eta}&	0.34626				&	0.22\times 10^{-3}	\\
\hline
\text{Total } \chi^2	&	\multicolumn{2}{c|}{69.5079}	\\
\hline
\end{array}
\end{math}
\end{center}
\caption{The best fit point values for the charged fermion masses (in GeV) and the Wolfenstein parameters of the CKM mixing matrix in the $\textbf{10} \oplus \textbf{120}$ model at the GUT scale.
The $x_i$ and $p_i$ columns correspond to the theoretical predicted values and the pulls.}
\label{bestfit10+120}
\end{table}

\subsection{The model with $\textbf{10} \oplus \textbf{120} \oplus \textbf{120}'$ Higgs sector}

The number of free real parameter of this model is 20 including
one parameter for the real coefficient $c_{10}$, 
four parameters for the complex coefficients $d_{120}^e$ and ${d'}_{120}^e$
three parameters for the real diagonal matrix $M_{10}$ (in an appropriate basis), 
twelve parameters for the complex antisymmetric matrices $M_{120}$, and $M'_{120}$.
The data fitting procedure for the $\textbf{10} \oplus \textbf{120} \oplus \textbf{120}'$ model results in the best fit point determined by
\begin{eqnarray}
c_{10}	&=&	66.335,	\\
d_{120}^e	 &=&	-0.93895 + 0.39900 i,	\\
{d'}_{120}^e &=&	-0.016416 -0.0047637 i,	\\
M_{10}	&=&
	\left(
	\begin{array}{ccc}
	0.00047530	&	0	&	0	\\
	0	&	0.050943	&	0	\\
	0	&	0	&	0.86104	\\
	\end{array}
	\right),	
\end{eqnarray}
\begin{eqnarray}
&& M_{120}		=	\nonumber \\	
&&	\left(
	\begin{array}{ccc}
	0 	&	0.0021268 -3.5349\times 10^{-5} i	&	-0.0081535  -0.00049252 i	\\
	-0.0021268 + 3.5349\times 10^{-5} i	&	0	&	-3.5746 \times 10^{-5} +0.16025 i	\\
	0.0081535  +0.00049252 i	&	3.5746 \times 10^{-5} -0.16025 i	&	0	\\
	\end{array}
	\right),	\nonumber \\
\\
&& M'_{120}	=	\nonumber \\
&&	\left(
	\begin{array}{ccc}
	0 	&	0.012290 +0.079333 i	&	0.19358 -0.11446 i	\\
	-0.012290 -0.079333 i	&	0	&	0.046671 + 13.5779 i	\\
	-0.19358 +0.11446 i	&	-0.046671 - 13.5779 i	&	0	\\
	\end{array}
	\right).	
\end{eqnarray}

In Table \ref{bestfit10+120+120}, the predicted values and the pulls for the observables of this model at the GUT scale are shown.
We see that there is no more tension on the second generation fermion masses ($m_c$, $m_s$) due to the model's flexibility with more degrees of freedom than that of the minimal one.
However, the tension in the third generation between the the pulls for the bottom and tau masses persists with smaller deviation for the tau mass:
\begin{eqnarray}
m_b^{\text{best-fit}} &=& \overline{m}_b + 2.4 \, \sigma_b ,	\\
m_\tau^{\text{best-fit}} &=& \overline{m}_\tau - 7.3 \, \sigma_\tau .
\end{eqnarray}

\begin{table}
\begin{center}
\begin{math}
\begin{array}{|c||c|c|}
\hline
\text{Observables}	&	x_i &	p_i	\\
\hline
\hline
m_u		&	0.39931\times 10^{-3}	&	0.17 \times 10^{-2}	\\
m_c		&	0.19919				&	7.7 \times 10^{-4}	\\
m_t		&	60.3537				&	0.10\times 10^{-2}	\\
m_d		&	0.86175\times 10^{-3}	&	0.042	\\
m_s		&	0.020710				&	0.50	\\
m_b		&	0.89166				&	2.4	\\
m_e		&	0.44049\times 10^{-3}	&	0.18 \times 10^{-2}	\\
m_\mu 	&	0.095107				&	0.42	\\
m_\tau	&	1.00705				&	7.3	\\
\lambda	&	0.22469				&	0.37\times 10^{-2}	\\
A		&	0.93248				&	0.17\times 10^{-2}	\\
\bar{\rho}&	0.13952				&	8.4 \times 10^{-7}	\\
\bar{\eta}&	0.34626 			&	3.2 \times 10^{-4}	\\
\hline
\text{Total } \chi^2	&	\multicolumn{2}{c|}{59.3001}	\\
\hline
\end{array}
\end{math}
\end{center}
\caption{The best fit point values for the charged fermion masses (in GeV) and the Wolfenstein parameters of the CKM mixing matrix in the 
$\textbf{10} \oplus \textbf{120} \oplus \textbf{120}'$ model at the GUT scale.
The $x_i$ and $p_i$ columns correspond to the theoretical predicted values and the pulls.}
\label{bestfit10+120+120}
\end{table}

\subsection{The model with $\textbf{10} \oplus \textbf{10}' \oplus \textbf{120}$ Higgs sector}

In this extended model, the number of free real parameter is 24 including 
one parameter for the real coefficient $c_{120}^u$, 
two parameters for the complex coefficient $c_{120}^e$, 
three parameters for the real diagonal matrix $M_{10}$ (in an appropriate basis),
twelve parameters for the complex symmetric matrix $M'_{10}$, and 
six parameters for the complex antisymmetric matrices $M_{120}$.
The best fit point for the $\textbf{10} \oplus \textbf{10}' \oplus \textbf{120}$ model is found to be
\begin{eqnarray}
c_{120}^u	&=&		-20.3343 ,	\\
c_{120}^e	&=&	
	3.08017  -5.24577\times 10^{-6} i,	\\
M_{10}	&=&
	\left(
	\begin{array}{ccc}
	-0.000717021 	&	0	&	0	\\
	0	&	0.0362958	&	0	\\
	0	&	0	&	0.875936	\\
	\end{array}
	\right),
\end{eqnarray}
\begin{eqnarray}
M'_{10}		&=&	
	\left(
	\begin{array}{ccc}
	0.0059128 +0.0054818 i 	&	-0.04779 +0.013064 i	&	0.11143 +0.13764 i	\\
	-0.04779 +0.013064 i	&	-0.082199 + 0.20885 i	&	4.37003 -0.61857 i	\\
	0.11143 +0.13764 i	&	4.37003 -0.61857 i	&	-34.4169 -49.0727 i	\\
	\end{array}
	\right),
\end{eqnarray}
\begin{eqnarray}
&& M_{120}		=	\nonumber \\
&&	\left(
	\begin{array}{ccc}
	0 	&	1.5193\times 10^{-5} -0.0013945 i	&	5.0427\times 10^{-5} +0.0053721 i	\\
	-1.5193\times 10^{-5} +0.0013945 i	&	0	&	9.7925\times 10^{-7}	-0.11571 i \\
	-5.0427\times 10^{-5} -0.0053721 i	&	-9.7925\times 10^{-7}	+0.11571 i	&	0	\\
	\end{array}
	\right)	.	\nonumber	\\
\end{eqnarray}

In Table \ref{bestfit10+10+120}, we present the predicted values and the pulls for the observables of this model at the GUT scale for the best fit point of this model.
Here, although the pull for the bottom mass is slightly smaller than that in the previous model due to a larger number of degrees of freedom, the large pull for the tau mass is still approximately the same:
\begin{eqnarray}
m_b^{\text{best-fit}} &=& \overline{m}_b + 2.4 \, \sigma_b ,	\\
m_\tau^{\text{best-fit}} &=& \overline{m}_\tau - 7.3 \, \sigma_\tau .
\end{eqnarray}
Meanwhile, as expected the fittings for other observables are much improved thanks to the large number of degrees of freedom in this model.

\begin{table}
\begin{center}
\begin{math}
\begin{array}{|c||c|c|}
\hline
\text{Observables}	&	x_i &	p_i	\\
\hline
\hline
m_u		&	0.398968\times 10^{-3}	&	2.4\times 10^{-5}	\\
m_c		&	0.199171				&	5.8\times 10^{-5}	\\
m_t		&	60.3555				&	3.9\times 10^{-6}	\\
m_d		&	0.861726\times 10^{-3}	&	0.042	\\
m_s		&	0.0207547				&	0.51	\\
m_b		&	0.891622			&	2.4	\\
m_e		&	0.440494\times 10^{-3}	&	0.20\times 10^{-2}	\\
m_\mu 	&	0.0951064				&	0.42	\\
m_\tau	&	1.00706				&	7.3	\\
\lambda	&	0.224686				&	2.7\times 10^{-4}	\\
A		&	0.932457				&	3.2\times 10^{-5}	\\
\bar{\rho}&	0.139518				&	5.2 \times 10^{-5}  \\
\bar{\eta}&	0.34626				&	4.7 \times 10^{-5}	\\
\hline
\text{Total } \chi^2	&	\multicolumn{2}{c|}{59.2984}	\\
\hline
\end{array}
\end{math}
\end{center}
\caption{The best fit point values for the charged fermion masses (in GeV) and the Wolfenstein parameters of the CKM mixing matrix in the 
$\textbf{10} \oplus \textbf{10}' \oplus \textbf{120}$ model at the GUT scale.
The $x_i$ and $p_i$ columns correspond to the theoretical predicted values and the pulls.}
\label{bestfit10+10+120}
\end{table}

In the alternative minimal model with $\textbf{10} \oplus \textbf{120}$ Higgs, both $M_{10}$ and $M_{120}$ that are used to fit down-type quark and charged lepton masses are also involved in the fitting u-type quark masses. Therefore, the tension is severe leading to a larger total $\chi^2$ value of 69.5079 for the best fit point.
In the extended models, one of these two matrices ($M_{120}$ in the $\textbf{10} \oplus \textbf{120} \oplus \textbf{120}'$ model, 
$M'_{10}$ in the $\textbf{10} \oplus \textbf{10}' \oplus \textbf{120}$ model)
is relaxed and does not involve in fitting u-type quark masses.
Therefore, the mild pull for $m_s$ of the minimal model 
(Eq. (\ref{pullms})) disappear, 
the tension between the pulls for $m_b$ and $m_\tau$ is slightly reduced, 
and other fittings are much improved 
in these two extension.
As a consequence, the best total $\chi^2$ values are 59.3001 and 59.2984 in the two extended models  respectively.
In all these considered models, there is always a tension between the pulls for bottom and tau masses.
It is due to the fact that down-type quark and charged lepton masses ($M_d$, $M_e$) in these three models are determined mainly by the symmetric matrix $M_{10}$ while the antisymmetric matrices ($M_{120}$, $M'_{120}$) are not enough to generate the adequate corrections.
Thus, large pulls of opposite sides persist for all three models.
Because of this property, even if more additional $\textbf{10}$-plets or $\textbf{120}$-plets Higgs fields of the same kind are introduced, the situation will not be improved.
%


\section{The minimal alternative model with bottom and vector-like quark mixing}

\subsection{Bottom and vector-like quark mixing}

From the results of data fitting to the alternative models, we can see the tension in the fitting of the bottom and tau masses.
In this section we consider a Yukawa interaction between the 3rd generation fermions, $\textbf{16}_3$, and the $\textbf{16}_D$ representation introduced in the minimal alternative model%
\footnote{
In general, the vector-like quarks have Yukawa interactions with all three generations of the SM quarks. In this analysis, we assume, for simplicity, that the dominant term is the Yukawa coupling with the third generation while other terms are negligible. This hierarchy is inspired from that in the SM. 
Such hierarchical picture should be governed by some underlying symmetry and deserves further study in the future.
}:
\begin{eqnarray}
\mathcal{L}_{SO(10)} \supset 
	Y_m \textbf{16}_3 \textbf{10}_H \textbf{16}_D . 
\label{mix}
\end{eqnarray}
Assuming that this Yukawa coupling is adequately small, we can neglect its RGE effects below the GUT scale.
After the $SO(10)$ symmetry breaking at the GUT scale, the new interaction in Eq. (\ref{mix}) results in the mixing mass term between the bottom quark and the vector-like quark $D_L$
while other heavy states are neglected:
\begin{eqnarray}
\mathcal{L}_{effective} \supset
	Y_m \alpha^1_d v_d \bar{b}_R D_L .
\end{eqnarray}
Due to such mixing, the bottom mass is deviated from that in the non-mixing case by a small amount while other charged fermion masses remain intact.

We parameterize the GUT mass relation for down-type quarks as follows:
\begin{eqnarray}
M_d	&=&	M_{10} + M_{120} + M_{mix},
\end{eqnarray}
where
\begin{eqnarray}
M_{mix}	&=&
	\left(
	\begin{array}{ccc}
	0	&	0	&	0	\\
	0	&	0	&	0	\\
	0	&	0	&	m_{vl}	\\
	\end{array}
	\right)
\end{eqnarray}
is the contribution from the above mass mixing with the vector-like quark.

\subsection{Data fitting}

Considering the mixing between the bottom and the vector-like quark $D_L$, we have performed the fitting to the charged fermion masses and mixing.
Relevant to this sector, this scenario has 16 free real parameters including 14 free inputs as those in the above minimal alternative model and 2 additional free inputs from the complex number $m_{vl}$.
The best fit point in this case is found to be

\begin{eqnarray}
c_{10}		&=&	40.3765 ,	\\
c_{120}^u	&=&	-31.5752 -40.1781 i ,	\\
c_{120}^e	&=&	-3.0042 -2.67699 i ,	
\end{eqnarray}
\begin{eqnarray}
M_{10}	&=&
	\left(
	\begin{array}{ccc}
	-0.000557118	&	0	&	0	\\
	0	&	0.00866987	&	0	\\
	0	&	0	&	-1.48365	
	\end{array}
	\right) ,
\end{eqnarray}
\begin{eqnarray}
M_{120}		&=&
	\left(
	\begin{array}{ccc}
	0	&	0.00032580 -0.0016479 i	&	0.0049921 -0.023637 i	\\
	-0.00032580 +0.0016479 i	&	0	&	0.054465 -0.078530 i	\\
	-0.0049921 +0.023637 i	&	-0.054465 +0.078530 i	&	0	\\
	\end{array}
	\right)	, \nonumber	\\
\end{eqnarray}
\begin{eqnarray}
m_{vl}	&=&	2.16619 -0.450918 i .	
\end{eqnarray}
Note that, for the above best fit point, 
$|m_{vl}| \ll m_t$ implies that the condition 
$|Y_m| \ll Y_t$ is satisfied (given that $|\alpha_1^d| v_d = \mathcal{O}(100)$ GeV).
This ensures the consistency of our assumption that the RGE effects of $Y_m$ on the considered observables are negligible.

In Table \ref{bestfit10+120mix}, the best fit values of the charged fermion masses and the Wolfenstein parameters of the CKM matrix are shown together with the corresponding pulls.
In this case, we have obtained a very good value for the total $\chi^2$ that is 0.217.
Therefore, the tension in fitting between the bottom and tau masses is resolved.

\begin{table}[h]
\begin{center}
\begin{math}
\begin{array}{|c||c|c|}
\hline
\text{Observables}	&	x_i &	p_i	\\
\hline
\hline
m_u		&	0.400616\times 10^{-3}	&	8.2\times 10^{-3}	\\
m_c		&	0.201931				&	0.089	\\
m_t		&	60.3277				&	0.016	\\
m_d		&	0.747056\times 10^{-3}	&	0.30	\\
m_s		&	0.016069				&	0.34	\\
m_b		&	0.829718				&	0.017	\\
m_e		&	0.44133\times 10^{-3}	&	0.040	\\
m_\mu 	&	0.0931455				&	2.0\times 10^{-3}	\\
m_\tau	&	1.58347				&	8.9\times 10^{-3}	\\
\lambda	&	0.224687				&	3.0\times 10^{-3}	\\
A		&	0.932497				&	3.3\times 10^{-3}	\\
\bar{\rho}&	0.139571				&	3.3\times 10^{-3}  \\
\bar{\eta}&	0.346291				&	3.1 \times 10^{-3}	\\
\hline
\text{Total } \chi^2	&	\multicolumn{2}{c|}{0.216997}	\\
\hline
\end{array}
\end{math}
\end{center}
\caption{The best fit point values for the charged fermion masses (in GeV) and the Wolfenstein parameters of the CKM mixing matrix in the 
$\textbf{10} \oplus \textbf{120}$ model with the Yukawa interaction between the 3rd generation and the vector-like quark $D$ (see Eq. (\ref{mix})) at the GUT scale.
The $x_i$ and $p_i$ columns correspond to the theoretical predicted values and the pulls.}
\label{bestfit10+120mix}
\end{table}

With the good fitting result for the charged fermion sector, it is necessary to verify if the model is capable to predict the realistic neutrino mass scale as obtained from cosmological observation 
\cite{Palanque-Delabrouille:2015pga}.
We find that the Dirac neutrino mass scale given by Eq. (\ref{mD1}) 
is of $\mathcal{O}(m_t)$ or larger for any value of the coefficient $c^n_{120}$.
On the other hand, the Majorana RH-neutrino mass scale is suppressed by a 2-loop factor of about $10^{-5} - 10^{-6}$ compared to the GUT scale due to the Witten mechanism.
Therefore, it is impossible to get the correct mass scale of the left handed neutrinos of $\mathcal{O}(0.1)$ eV 
\cite{Palanque-Delabrouille:2015pga} 
by the seesaw type I mechanism.

\subsection{Two-step symmetry breaking}

To ameliorate the problem that seesaw scale through the Witten mechanism is too low, we consider 
that the GUT symmetry breaking chain occurs at two distinct scales with $SO(10) \rightarrow SU(5)$ at $M_{SO(10)}$, and
$SU(5) \rightarrow SU(3)_C \times SU(2)_L \times U(1)_Y$ at $M_G = 1.5
\times 10^{16}$ GeV $<M_{SO(10)}$.
The $SO(10)$ GUT symmetry breaking scale is a free parameter bounded from above by the (reduced) Planck scale, 
$M_{SO(10)} \lesssim M_\text{Planck} = 2.4 \times 10^{18}$ GeV.
In this case, the RH neutrino Majorana mass through the Witten mechanism is found to be
\bea
M_R = \left( \frac{m_q}{M_W} \right) \epsilon_{10} \left( \frac{\alpha_G}{\pi} \right)^2 M_{SO(10)}.
\label{MR2}
\eea
Note that for 
$M_{SO(10)} \sim M_\text{Planck}$, the RH neutrino scale becomes of $\mathcal{O}(10^{13})$ GeV, resulting in 
an acceptable LH neutrino mass scale of about $\mathcal{O}(0.1)$ eV.

We repeat the data fitting analysis to the masses and mixings of all fermions including those of the neutrinos, 
but with $M_{SO(10)}$ being a new free parameter.
Since $M_G$ is close to $M_{SO(10)}$, 
we have neglected the RGE evolutions of the fermion masses and mixings from $M_G$ to $M_{SO(10)}$,
and have used the $SO(10)$ GUT mass relations, Eqs. (\ref{mu1})-(\ref{me1}), to match with experimental data at $M_G$.
In our analysis, the radiatively generated RH-neutrino Majorana mass is parameterized as follows
\begin{eqnarray}
M_R &=& c_{10}^R M_{10} .
\end{eqnarray}
We have obtained the best fit point with the $\chi^2$ value of 7.8198 that is experimentally viable.
The parameters corresponding to this point are determined as

\begin{eqnarray}
c_{10}		&=&	40.6978 ,	\\
c_{120}^u	&=&	-17.6178 -10.7482 i ,	\\
c_{120}^e	&=&	-3.151 + 6.33661 i ,	\\
c_{120}^n	&=&	47.9034 -38.7087 i ,	\\
c_{10}^R	&=&	-3.81633\times 10^{12} + 3.22151\times 10^{11} i,
\end{eqnarray}
\begin{eqnarray}
M_{10}	&=&
	\left(
	\begin{array}{ccc}
	-3.9037\times 10^{-5}	&	0	&	0	\\
	0	&	0.0045238	&	0	\\
	0	&	0	&	-1.48135	
	\end{array}
	\right) ,
\end{eqnarray}
\begin{eqnarray}
M_{120}		&=&
	\left(
	\begin{array}{ccc}
	0	&	0.0007942 -0.0005567 i	&	0.001138 -1.6085\times 10^{-6} i	\\
	-0.0007942 +0.0005567 i	&	0	&	0.042495 +0.034654 i	\\
	-0.001138 +1.6085\times 10^{-6} i	&	 -0.042495 -0.034654 i	&	0	\\
	\end{array}
	\right)	,	\nonumber \\
\end{eqnarray}
\begin{eqnarray}
m_{vl}	&=&	2.20452 -0.402702 i .	
\end{eqnarray}
For this best fit point, the RH neutrino Majorana mass is
$M_R \approx 5.6 \times 10^{12}$ GeV.
Taking $\epsilon_{10} = 0.11$ as an example, the $SO(10)$ GUT breaking scale is found to be 
$M_{SO(10)} \approx 2.35 \times 10^{18}$ GeV, 
which is smaller than $M_\text{Planck}$, 
and we neglect gravity effects in our analysis.

The predicted observables and their pulls for the best fit point are given in Table \ref{bestfit10+120mix2}.
The largest deviation is 2.3$\sigma$ corresponding to the strange quark mass.
Therefore, a precise determination of the strange quark mass in the near future can test the validity of this model.

\begin{table}
\begin{center}
\begin{math}
\begin{array}{|c||c|c|}
\hline
\text{Observables}	&	x_i &	p_i	\\
\hline
\hline
m_u		&	0.412643\times 10^{-3}	&	0.068	\\
m_c		&	0.203016				&	0.12	\\
m_t		&	60.309				&	0.027	\\
m_d		&	0.20426\times 10^{-3}	&	1.5	\\
m_s		&	0.504091\times 10^{-2}	&	2.3	\\
m_b		&	0.831351				&	0.080	\\
m_e		&	0.441872\times 10^{-3}	&	0.065	\\
m_\mu 	&	0.0932523				&	0.021	\\
m_\tau	&	1.57706					&	0.072	\\
\lambda	&	0.224687				&	0.19\times 10^{-2}	\\
A		&	0.932724				&	0.022	\\
\bar{\rho}&		0.139633			&	0.71\times 10^{-2}  \\
\bar{\eta}&		0.346093			&	0.017	\\
\Delta m_{21}^2	&	1.1119\times 10^{-4}	&	0.78\times 10^{-2}	\\
\Delta m_{31}^2	&	3.79811\times 10^{-3}	&	0.35\times 10^{-2}	\\
\theta_{12}	&	0.573177	&	0.026	\\
\theta_{23}	&	0.866723	&	0.041	\\
\theta_{13}	&	0.150271	&	2.8\times 10^{-4}	\\
\delta_\text{CP}	&	-2.49584	&	2.9\times 10^{-5}	\\
\hline
\text{Total } \chi^2	&	\multicolumn{2}{c|}{7.8198}	\\
\hline
m_{\beta\beta}	&	\multicolumn{2}{c|}{0.22 \text{ meV}}	\\
\hline
\Sigma			&	\multicolumn{2}{c|}{0.078 \text{ eV}}	\\
\hline
\end{array}
\end{math}
\end{center}
\caption{The best fit point values  at $M_G$ for the 
charged fermion masses (in GeV), the Wolfenstein parameters of the CKM mixing matrix, 
neutrino mass squared differences (in eV$^2$), 
and the parameters of the PMNS mixing matrix in the two-step symmetry breaking scenario of the
$\textbf{10} \oplus \textbf{120}$ model with the Yukawa interaction between the 3rd generation and the vector-like quark $D$ (see Eq. (\ref{mix})).
The $x_i$ and $p_i$ columns correspond to the theoretical predicted values and the pulls.
In the last three rows, the total $\chi^2$ value, the predicted values of the effective Majorana neutrino mass, $m_{\beta\beta}$, and the sum of light neutrino masses, $\Sigma$, for the best fit point are presented.}
\label{bestfit10+120mix2}
\end{table}

For consistency, we take into account the current constraints on the neutrino sector from nuclear physics and astrophysics.
The Majorana nature of light neutrinos can be revealed by the neutrinoless double beta decay. The current constraint on the effective Majorana mass reads 
\cite{Alvis:2019sil}
\begin{eqnarray}
m_{\beta\beta}  \equiv 
	\left|
	\sum_{i=1}^3 m^\nu_i U_{ei}^2 e^{i\varphi_i}
	\right|		& \lesssim &
	\mathcal{O}(100) \text{ meV} ,
\end{eqnarray}
where $m^\nu_i, U_{ei}$, and $\varphi_i$ are the light neutrino mass eigenvalues, the PMNS matrix elements, and the Majorana phases, respectively.
The effective Majorana mass predicted by our best fit point is found to be  
$m_{\beta\beta} = 0.22$ meV, that is well below the current upper limit.
On the other hand, the sum of light neutrino masses is constrained by the measurement of the  Cosmic Microwave Background (CMB) anisotropy
\cite{Palanque-Delabrouille:2015pga}:
\begin{eqnarray}
\Sigma = m^\nu_1 + m^\nu_2 + m^\nu_3
	& \lesssim &
	0.12 \text{ eV}.
\end{eqnarray}
For the best fit point in our scenario, $\Sigma$ is found to be 0.078 eV, which is consistent with the upper bound.

\section{Conclusion}

In the alternative $SO(10)$ GUTs, only $\textbf{10}$-plets and $\textbf{120}$-plets of Higgs fields are involved in generating masses for the SM charged fermions.
The RH-neutrinos with vanishing Majorana masses at tree level obtain nonzero ones at quantum two-loop level via the Witten mechanism.
The LH-neutrinos acquire tiny masses in charge of neutrino oscillations via the seesaw mechanism.
The success of the gauge coupling unification is achieved by introducing the vector-like quarks with the mass scale of $\mathcal{O}$(TeV).
The existence of these vector-like fermions deflects the SM RGE trajectories, resulting in a unique pattern of fermion masses and mixing at the GUT scale.
The matching between such a particular effective model at low energies and the alternative $SO(10)$ models has been investigated for the first time in this paper.
The experimental data for the charged fermion masses and the CKM matrix are evolved to the GUT scale and have been used to fit the model parameters.
In the alternative minimal model with $\textbf{10} \oplus \textbf{120}$ Higgs fields,
we have found the best fit point among the space of 14 free parameters. The total $\chi^2$ value of this point is 69.5079.
Large pulls are observed corresponding to $m_s$, $m_b$, and $m_\tau$.
In two simple extensions to the minimal models with 
$\textbf{10} \oplus \textbf{120} \oplus \textbf{120}'$ and
$\textbf{10} \oplus \textbf{10}' \oplus \textbf{120}$ of Higgs fields,
there are more degrees of freedom with 20 and 24 free parameters, respectively.
The results of data fitting are improved with 
the best $\chi^2$ equals 59.3001 for the former and 59.2984 for the latter.
However, due to the GUT mass relation among down-type quarks and charged leptons, the severe tension between the pulls for $m_b$ and $m_\tau$ still exists.
Especially, the tau mass always suffers from large pulls of about 7.3 that rule out all the three models.

We have shown that the tension in the fitting to the bottom and tau masses in the minimal alternative model can be resolved by introducing the new Yukawa interaction between the 3rd generation fermions, $\textbf{10}_3$, the vector-like fermions, $\textbf{10}_ D$, and the $\textbf{10}$-plet of Higgs fields.
Neglecting all the heavy fields, this interaction results in the mixing between the bottom quark and the vector-like quark $D$.
We have found that the best $\chi^2$ value in the data fitting to the charged fermion sector is 0.217.
However, the Dirac neutrino mass scale is too large in this scenario such that
the 2-loop suppressed Majorana RH-neutrino mass scale could not lead to the LH-neutrino mass scale compatible with its cosmological bound.
To ameliorate this problem of the radiatively generated seesaw scale being too small, 
we have considered the two-step symmetry breaking scenario.
Here, the GUT symmetry breaking chain occurs at two distinct scales with $SO(10) \rightarrow SU(5)$ at $M_{SO(10)}$, and 
$SU(5) \rightarrow SU(3)_C \times SU(2)_L \times U(1)_Y$ at $M_G = 1.5 \times 10^{16}$ GeV.
Performing the data fitting analysis in this case, we have found the best fit point with $\chi^2_\text{best} = 7.8198$ that is experimentally allowed.
For this point, the largest pull of 2.3 comes from the strange quark mass.
Thus, a more precise determination of the strange quark mass can test this scenario.
We have found the effective Majorana neutrino mass of $m_{\beta\beta} = 0.22$ meV and the sum of light neutrino masses as $\Sigma = 0.078$ eV, which are consistent with the current constraints from the search for the neutrinoless double beta decay and the CMB anisotropy measurement, respectively.

To summarize, in this work, almost all possibilities have been exhausted for the alternative models with $\textbf{10}_H$ and $\textbf{120}_H$. 
The vector-like quarks turn out to be important not only for the success of the gauge coupling unification, but also for the fitting of charged fermion masses and mixing.
In the context of the Witten mechanism, the two-step symmetry breaking is necessary for the prediction of observables in both charged and neutral fermion sectors consistent with experimental results.

\section*{Acknowledgments}

This work is supported in part by JSPS Kakenhi No.~17H01133 (T.F.), the U.S. Department of Energy No.~DE-SE0012447 (N.O.), and Vietnam National Foundation for Science and Technology Development (NAFOSTED) No.~103.01-2017.301 (H.M.T.).

\end{document}